# SafeVchat: Detecting Obscene Content and Misbehaving Users in Online Video Chat Services

Xinyu Xing[1], Yu-Li Liang[1], Hanqiang Cheng[2], Jianxun Dang[2], Sui Huang[3],Richard Han[1], Xue Liu[2], Qin Lv[1], Shivakant Mishra[1]
[1]University of Colorado at Boulder, [2]McGill University, [3]Ohio State University

Department of Computer Science
University of Colorado at Boulder

# SafeVchat: Detecting Obscene Content and Misbehaving Users in Online Video Chat Services

Xinyu Xing[1], Yu-Li Liang[1], Hanqiang Cheng[2], Jianxun Dang[2], Sui Huang[3], Richard Han[1], Xue Liu[2], Qin Lv[1], Shivakant Mishra[1]

[1]University of Colorado at Boulder, [2]McGill University, [3]Ohio State University
{xinyu.xing, yu-li.liang}@colorado.edu (co-first authors)

## ABSTRACT

Online video chat services such as Chatroulette, Omegle, and vChatter that randomly match pairs of users in video chat sessions are fast becoming very popular, with over a million users per month in the case of Chatroulette. A key problem encountered in such systems is the presence of flashers and obscene content. This problem is especially acute given the presence of underage minors in such systems. This paper presents SafeVchat, a novel solution to the problem of flasher detection that employs an array of image detection algorithms. A key contribution of the paper concerns how the results of the individual detectors are fused together into an overall decision classifying the user as misbehaving or not, based on Dempster-Shafer Theory. The paper introduces a novel, motion-based skin detection method that achieves significantly higher recall and better precision. The proposed methods have been evaluated over real-world data and image traces obtained from Chatroulette.com.

## 1. INTRODUCTION

Online video chat services have become increasingly popular, and include such systems as Chatroulette [9], BlurryPeople [5], RandomDorm [34], Omegle [31], vChatter [24], etc. While most of these services have been introduced only recently (less than one year ago), statistics show that membership in for example Chatroulette has grown by 500% since the beginning of 2010. Furthermore, tens of thousands of users are online in Chatroulette at any point of time, 24 hours a day. In July 2010, 1.3 million US users [22] and 6.3 million users in total [39] are estimated to have visited Chatroulette.

A common feature of such online video chat services is that they randomly match pairs of users via video connections. In the most popular cases, users are anonymous from each other, and need not supply any overt information to the system or other users that may identify them, such as their name or profile. Users of such systems can quickly move on to the next random pairing by clicking “Next”. In addition, such services are typically offered for free, and are easy to use, which enhances their popularity.

A key problem encountered in such anonymous video chat systems is that many of the users may engage in sexually explicit behavior such as flashing/revealing themselves with full frontal nudity. This misbehavior drives away other users that are not interested in viewing such obscene content, thus limiting such systems from achieving their full popularity. In addition, flasher-type misbehavior has potential legal ramifications since many of the viewers attracted to such video chat systems are minors [14]. Our observations on a typical Saturday night indicate that as many as 20-30% of Chatroulette users are minors. Even though users must confirm that they are at least 16 years old and agree not to broadcast obscene, offending or pornographic material, it is nearly impossible to enforce this due to the lack of login or registration requirements of such anonymous systems.

One approach to deal with misbehavior is to de-anonymize users, on the hypothesis that requiring users to reveal their identities will then force users to behave. For example, vchatter uses such an approach on Facebook, where only Facebook users may participate in its Facebook application, and the Facebook name of users in the video chat are displayed during each video pairing session. This approach of using de-anonymization to address misbehaving users has several drawbacks. First, it undercuts one of the primary motivations that attracts users to the most popular roulette-type video services, namely that people can participate anonymously for fun. Second, flashers can circumvent this approach by registering false/dummy profiles.

An alternative approach to dealing with misbehavior is to analyze the video sessions to detect obscene content, and thus flag misbehaving users. Some video chat sites have taken manual crowd-sourcing to detect misbehaving users, i.e. images/screen captures are sent to real humans who are paid to manually identify whether the user is a flasher or not [9]. Such approaches incur a high economic cost and thus are not scalable. Moreover, such approaches are not necessarily applied uniformly, i.e. only images that are ”reported” (Chatroulette for example has a ”Report” button to flag misbehaving users) are actually inspected. Our analysis indicates that not all misbehaving users are reported, and some users who are not misbehaving are still reported as such by pranksters.

A more promising and scalable approach is automated detection of misbehaving users, by employing image recognition algorithms. In particular, we investigate a novel solution to the problem of flasher detection that fuses together an array of evidences that image recognition algorithms refine. To the best of our knowledge, this is the first work investigating this problem space. This paper claims the following key contributions:

- we identify and analyze the issues introduced by online video chat systems, present statistical studies of its users, and demonstrate that existing pornography detection algorithms don’t work well in this problem space.

- we describe a holistic solution to detect flashers and obscene content that (1) adapts this problem space to a set of evidence that image recognition algorithms refine, including evidence of a face, eyes and nose, etc.,

(2) introduce our own novel motion-based skin detector to supplement classification of misbehaving users, and (3) we *fuse* these individual evidences by building a probabilistic model using Dempster-Shafer Theory (DST) [36] to obtain a joint classification of each user as misbehaving or not.

- we evaluate the effectiveness of our fused classifier system in terms of its precision and recall by employing real world data sets obtained from Chatroulette.com.

The research reported in this work is in close collaboration with Chatroulette. We thank Andrey Ternovskiy, the founder of Chatroulette.com, for providing us with these otherwise unobtainable internal data traces. The observations, algorithm design and system evaluations are carried out based on the dataset. It is worth noting that though we use the Chatroulette system as a running example in this paper, our approach is generally applicable to all roulette-type video chat services.

In the following, we describe related work, present a statistical analysis of users using the Chatroulette dataset, explain the limitations of existing techniques, and glean key observations from the user data. We then describe the architecture of our fused classifier system - SafeVChat, including the design requirements, and individual system components. Next, we describe our motion-based skin detection algorithm. We then detail the probabilistic model used to fuse together the results of the individual classifiers leveraging Dempster-Shafer Theory. Finally, we provide a rigorous evaluation of the effectiveness of our joint classification system using the data traces obtained from within Chatroulette.com.

## 2. RELATED WORK

Content-based image analysis has been an active research area for over a decade (see [12, 28] for some surveys). In particular, skin-color modeling and skin detection techniques have been successfully used for detecting pornographic content with high accuracy in Internet videos [25, 19, 27, 7, 6, 41, 35, 23, 21]. However, skin appearance may vary significantly for different illumination, background, camera characteristics, and ethnicity, which are particularly true in online video chat images. A recent survey study [26] concludes that skin detection methods may only be used as a preprocessor for obscene content detection, and other content types such as text [19] and motion analysis [21] may be incorporated to improve accuracy. In this paper, we propose a novel, motion-based skin detection method that analyzes several consecutive video frames. This method achieves much higher recall and better precision than PicBlock [32], a widely-used pornographic image blocker.

Besides skin detection, our work also leverages an array of image recognition techniques, including face, nose, eye, mouth, and upper body detectors, which are supported in the latest version of OpenCV library [20]. An extensive survey of these techniques is outside the scope of this paper. We focus on adapting these techniques to our online video chat images and identifying the strengths and limitations of each individual technique. Such information is then utilized in the fusion process.

A number of (ensemble) methods have been proposed to combine or fuse multiple classifiers in order to reach a final classification decision. Two representative methods are Dempster-Shafer Theory (DST) [36] and AdaBoost [15]. Compared to AdaBoost, DST supports multi-level decisions, and further considers the reliability of the evidence provided by

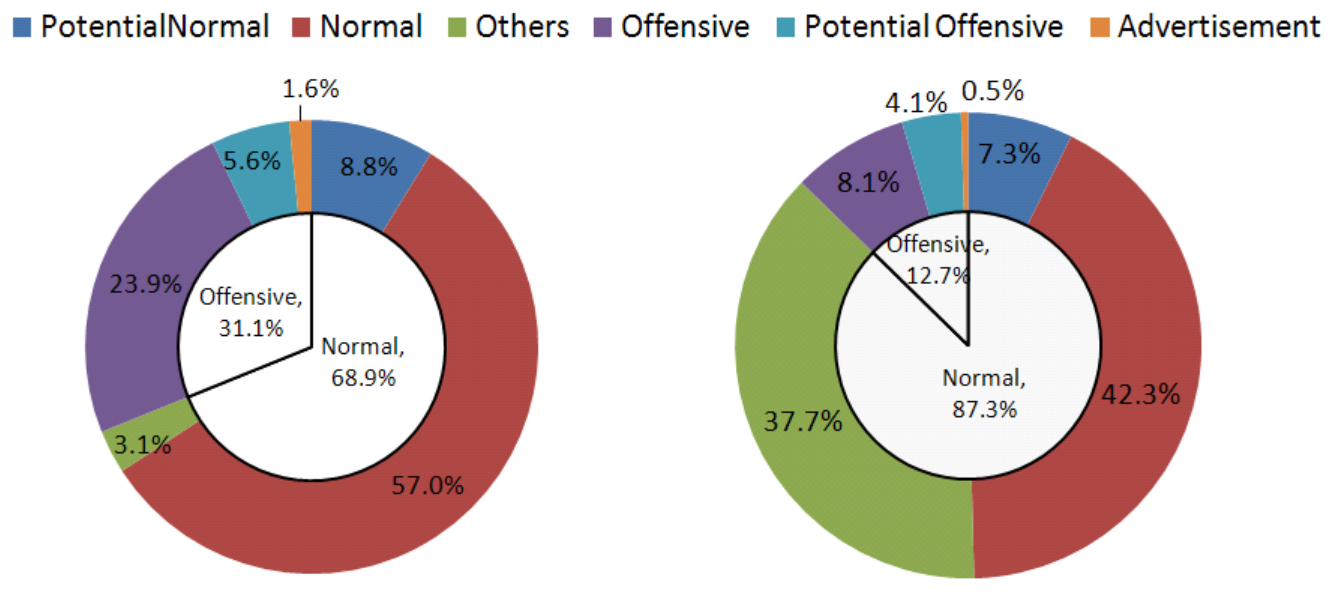

**(a) Original dataset** **(b) Filtered dataset**

**Figure 1: Categorization and percentage of different types of users in the Chatroulette dataset.**

each classifier. DST has been applied in a variety of classification settings, such as [29, 3, 10, 30]. The key challenge of using DST is to define the appropriate mass function (or basic probability) assignment, which is dependent on specific applications. One of the main contributions of this work is how we design mass function and apply DST to detecting obscene content in online video chat services.

## 3. DATA ANALYSIS AND OBSERVATIONS

In this section, we analyze a real-world dataset obtained from Chatroulette. After an initial statistical analysis of this dataset, we then investigate a state-of-the-art commercial software for pornography detection. We identify the limitations of this technique and present the key observations derived from our dataset with regard to obscene content and misbehaving users.

### 3.1 Dataset and Statistics

We have obtained a real-life dataset, provided by the founder of Chatroulette Andrey Ternovskiy, containing screenshot images from 20,000 users. Each screenshot image has a 320 × 240 resolution. Based on our analysis of this dataset, we categorized these 20,000 users into two types: offensive and normal users. *Offensive or misbehaving users* are defined based on the following criteria. If a user broadcasts obscene content or behaviors (*offensive content*), or intentionally shows his/her naked chest without face in front of the webcam (*potential offensive content*), or broadcasts pornographic advertisements (*advertisement content*), then the user is defined as an offensive user. On the contrary, *normal users* are chatters who stay fully clothed. A majority of normal users show their faces in front of webcams (we call the content that these users broadcast as *normal content*) and some others point their webcams to their clothed chest (we call the content that these users broadcast as *potential normal content*). In addition to these two types of content, there is a special type of normal users who completely block their webcams, or point their webcams to a static scene (e.g. a room corner or ceiling), or broadcast pre-prepared interesting videos etc. (we call the content that these normal users broadcast as *other content*). Figure 1 (*original dataset*) shows the categorization and distribution of different types of users and content in the Chatroulette dataset. Since Chatroulette took screenshots for all the users at approximately the same time and randomly selected 20,000 users as well as their screenshot images, we believe the Chatroulette dataset is representative of Chatroulette users' characteristics.

### 3.2 Limitations of Exisitng Techniques

Existing techniques for detecting objectionable content are mostly content-based, which typically utilize both image recognition and text analysis techniques to filter out pornographic webpages. Different from detecting pornographic webpages, detecting objectionable content in the context of online video chat systems cannot rely on text message analysis, since the content communicated between chatters does not pass through the central server and obtaining text messages that were exchanged is not practical. Furthermore, some chatters may have a conversation using audio devices instead of typing messages. Consequently, we focus on investigating whether state-of-the-art image recognition techniques, which are used for pornographic website detection, can be applied to objectionable content detection in the context of online video chat systems.

Specifically, we used PicBlock 4.2.3 (Pornographic Image Blocker), a state-of-the-art commercial software [32], to classify 1,000 user screenshot images which were randomly selected from the Chatroulette dataset. Surprisingly, even though PicBlock usually achieves high accuracy when detecting pornographic images on websites, it performed poorly on the screenshot images of online video chat users – the precision and recall for correctly detecting misbehaving users were only 0.253 and 0.390, respectively. This is mainly due to the large diversity in illumination, sensing quality of web cameras used by different chatters, ethnicity (African, Asian, Caucasian and Hispanic groups), and variations in individual user characteristics such as age, sex, what is prominently displayed in the image, and so on. Other factors such as appearances (makeup, hairstyle and glasses etc.), backlighting, shadows and motion also have significant influence on skin-color appearance. Indeed, these issues have also been investigated in a recent survey study [26], which concluded that the skin detection methodology can only be used as a preprocessor for obscene content detection.

## 3.3 Key Observations

Using the Chatroulette dataset, we have conducted further analysis and identified some discriminative characteristics that are specific in online video chat services. (1) Misbehaving users on online video chat systems usually hide their faces because they are not willing to compromise their identity privacy. (2) Different from regular pornographic images, misbehaving users may not completely expose themselves. For example, some misbehaving users may only expose their genitals in front of the webcam and stay partially clothed. (3) Chatters who present their faces in front of webcams are mostly normal users because showing both the body trunk and the face of a user requires the user placing his/her webcam far from the user, but chatters who do not show their faces may not be flashers. (4) A fair amount of chatters do not show their faces clearly. For example, only a partial face is presented in front of the webcam. (5) Webcams are usually set up in a stable way, i.e., a majority of chatters do not keep moving or adjusting their webcams. In the following sections, we take advantage of these observations to design our obscene content detection system.

# 4. ARCHITECTURE

The primary goal of our system is to detect users who abuse online video chat services, namely those who display obscene content. In this section, we first discuss the system design requirements and assumptions about Chatroulette's capabilities in terms of providing our system with data for analysis purposes. Following that, we describe the architecture of our SafeVchat system.

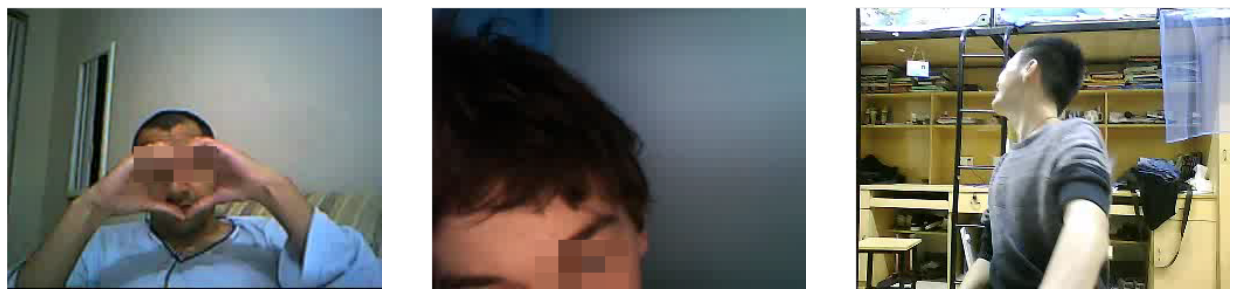

(a) (b) (c)

**Figure 2: A few screenshot images from Chatroulette.**

## 4.1 Design Requirements

To design our obscene content detection system, there are several important requirements. First, in terms of correctly classifying a misbehaving user, the precision should be high because in a system such as Chatroulette, all users classified as misbehaving will be subject to further costly review by crowdsourcing. We therefore want to be precise in only classifying users as misbehaving who are truly misbehaving, as any false positives incur an additional economic cost. We also want our recall to be high, since if we miss a misbehaving user via a false negative, these users will appear in the chat system. The resulting cost is twofold: adult users will be put off by the obscene content and may leave the system; further, minors would be subject to inappropriate content.

In an online video chat system, for system scalability, the video stream is transmitted in a peer-to-peer manner after a pair of users have been matched by the server. It is prohibitively expensive to obtain and monitor users' complete video content from a centralized server. To detect offensive content, the Chatroulette server is however able to partially obtain a user's video content by periodically taking sequential screenshot images from the user. The sequential screenshot images are taken at a predefined fixed time interval. Currently, in one period, three screenshot images are taken. Therefore, our obscene content detection system is designed based on the assumption that systems like Chatroulette are capable of providing this capacity.

## 4.2 System Architecture

Our key approach is to use evidence from multiple classifiers to strengthen the overall classification of whether a user is misbehaving or not. We consider the following types of evidence while designing our detection system, namely evidence of the presence of face, eyes, nose and mouth, upper body evidence and skin exposure evidence. Many of the individual classifiers that we will use are based on detecting a human face, since our observations from Section 3 indicate that misbehaving users hide their faces, while chatters who reveal their faces are typically normal users. However, a face detector needs to be augmented by additional facial evidences, since many of the following scenarios may occur based on our observations:

- In the process of taking a screenshot for a user, the user's face might be blocked by the user's hands. Thus this face evidence alone is not able to be refined by our system. (See Figure 2(a)).
- A user may intentionally show his face partially and our system can only refine partial facial evidences. For example, Figure 2(b) shows that only eyes are present in the screenshot image.
- Even though we use the state-of-the-art face detector to refine face evidence, there is still the possibility that

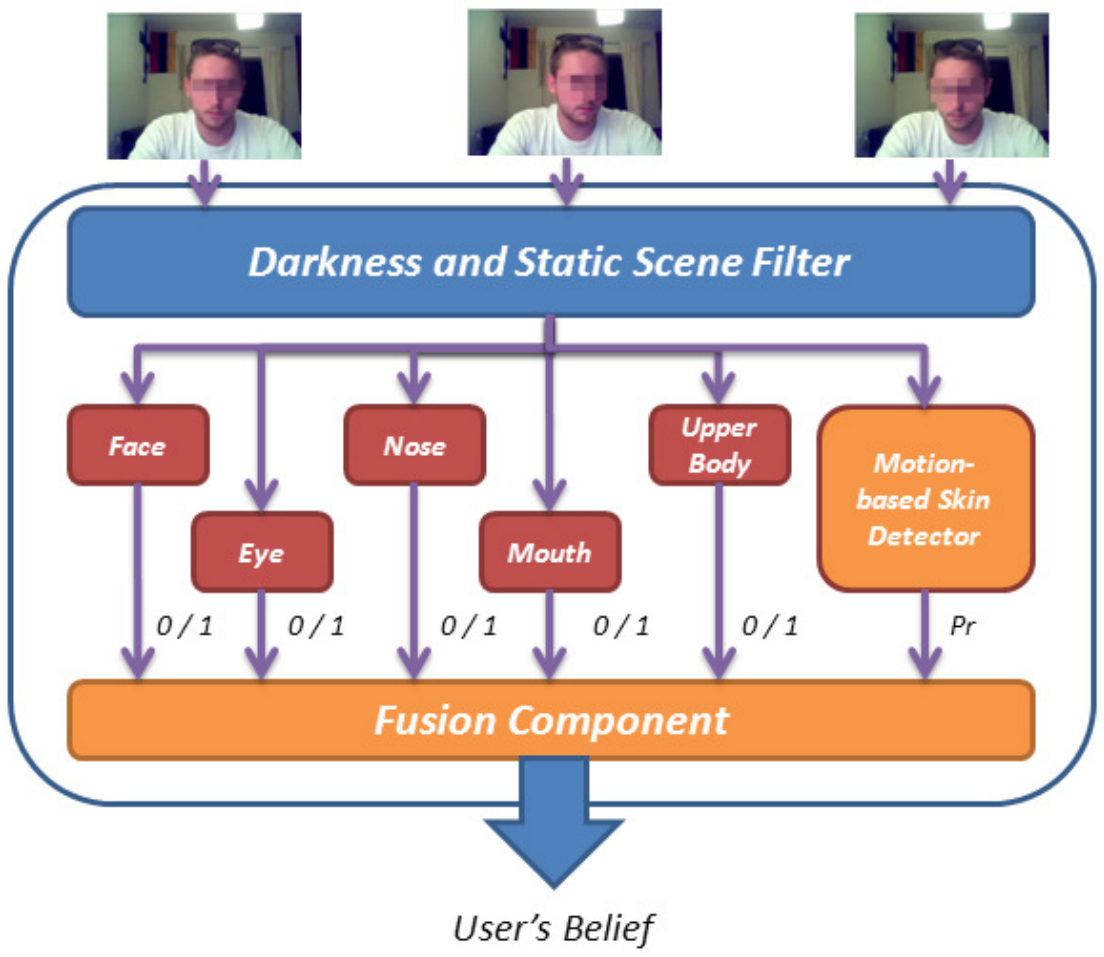

**Figure 3: SafeVchat: System architecture for detecting misbehaving users in online video chat services.**

the face cannot be detected (See Figure 2(c)).

Figure 3 shows the architecture of our SafeVchat system. The system contains five main components including a darkness filter, facial detectors, an upper body detector, a skin exposure detector and a fusion component. The darkness filter is used for identifying users who operate their webcams in the dark. The remaining users are subject to fused classification. Facial detectors contain a set of individual classifiers - face, eye, nose and mouth classifiers. The upper body detector is used for identifying whether an upper body is present. The classifiers that our system uses for facial and upper-body detection are provided by the OpenCV library and their outputs are all binary values to indicate presence or absence. The motion-based skin detector is a motion and skin color based classifier which we designed for determining the probability that a user is a misbehaving chatter. Finally, the fusion component of SafeVchat is used for combining evidences from facial detectors, the upper body detector and the motion-based skin detector to refine and make a final probabilistic decision. The fusion component is based on Dempster-Shafer Theory, and will be discussed in detail in Section 6.

The work flow of our system is as follows: When the three sequential screenshot images of a user are fed into our detection system from Chatroulette, the darkness detector first determines whether the user is using his or her webcam in the dark. If yes, a warning notification will be sent to the user. Otherwise, these screenshot images will be passed to the facial detectors, upper body detector and skin exposure detector. These detectors use their respective classifiers to refine evidence from the three screenshot images and pass their refined evidences to the fusion component. The fusion component harnesses the refined evidences to make a probabilistic decision and output how likely the user is classified as a misbehaving user.

# 5. MOTION-BASED SKIN DETECTION

We mentioned earlier that existing color-based skin detection techniques do not work well for the screenshot images of online video chat users. Specifically, we tested two existing methods: the *Adaptive real-time skin detector* from OpenCV [11, 20], and *PicBlock*, a commercial software for pornography detection. Neither method performed well on our video chat dataset (see Section 7 for details). This is mainly due to the diversity of skin color appearance (e.g., illumination, webcam characteristics) and skin-color like objects/backgrounds (e.g., yellowish sofa or white wall with yellowish lighting). As a result, real user skin may not be detected, while non-skin objects/backgrounds may be misclassified as skin leading to low flasher detection accuracy in online video chat services.

We also observe that in online video chat sessions, the moving parts in images are usually the region of interest – e.g., normal users moving their heads and clothed body parts, or flashers moving their naked body parts or touch naked parts with hand. Users with larger "non-face skin" exposure in such "target regions" are more likely to be flashers. The adaptive skin detector does consider motion in images, but its skin color palette does not capture the diversity of skin colors in online video chat images. In addition, it uses the optical flow method, which assumes only small motion in video frames. This assumption may not hold in online video chat services with large number of concurrent users – capturing high frequency video frames is infeasible and screenshots can only be captured with larger time intervals (e.g., 10 seconds in our dataset). Therefore, user movement between two consecutive screenshots can be significant.

In this work, we propose a novel motion-based skin detection method to detect obscene content and misbehaving users in online video chat systems. Our method consists of four major components: (1) calculates the *target map* (which contains the target region) via motion in consecutive screenshots; (2) uses a new *skin detector* with different skin color palettes to detect "non-face skin" in the target region; (3) calculates the *skin proportion*, which is the ratio of non-face skin area to target region; and (4) determines the *misbehaving probability* based on the skin proportion.

## 5.1 Target Map

Given two consecutive screenshots of a user, we define the *target map* to identify changes (motion). This is achieved through image subtraction, i.e., subtracting the pixel values at corresponding positions of the two images. For example, if a normal user moves his/her hand against the background wall, both the hand and the wall are included in the target region by image subtraction, and the target region contains both skin (i.e. hand) and non-skin (i.e. wall). If a flasher touches his/her naked part with hand, the target region contains mostly skin (i.e. hand and naked parts). Therefore, via image subtraction, target map captures the region we are interested in for better detection accuracy.

To avoid the noise introduced by individual pixels, we calculate the difference between *tiles*, and each tile is a rectangle containing multiple pixels. Specifically, each image is divided into $N \times N$ tiles ($N$ is an integer). For each tile $T_{rc}$ ($r, c \in [1, N]$), let $x_i (i \in [1, n])$ be one of the $n$ pixels in that tile, we calculate the tile's average intensity of RGB channels as follows:

$$\overline{T}_{rc} = \Sigma_{i=1}^{n} (R_{xi} + G_{xi} + B_{xi})/(n \times 3), \qquad (1)$$

where $R_{x_i}, G_{x_i}, B_{x_i}$ are the R, G, B color channel intensities of pixel $x_i$, respectively. Therefore, the target map of any two given images contains $N \times N$ elements, each representing the difference between two corresponding tiles' average RGB intensities. An element is set to 1 if the absolute difference is above the threshold, and 0 otherwise. The threshold (set to 9 in our experiemts) is determined based on the average difference of manually picked static images (i.e., no movement). Target maps are further improved by filling holes and removing glitches via morphological filter (e.g., erosion and dilation operation) [13]. See examples in Figure 4.

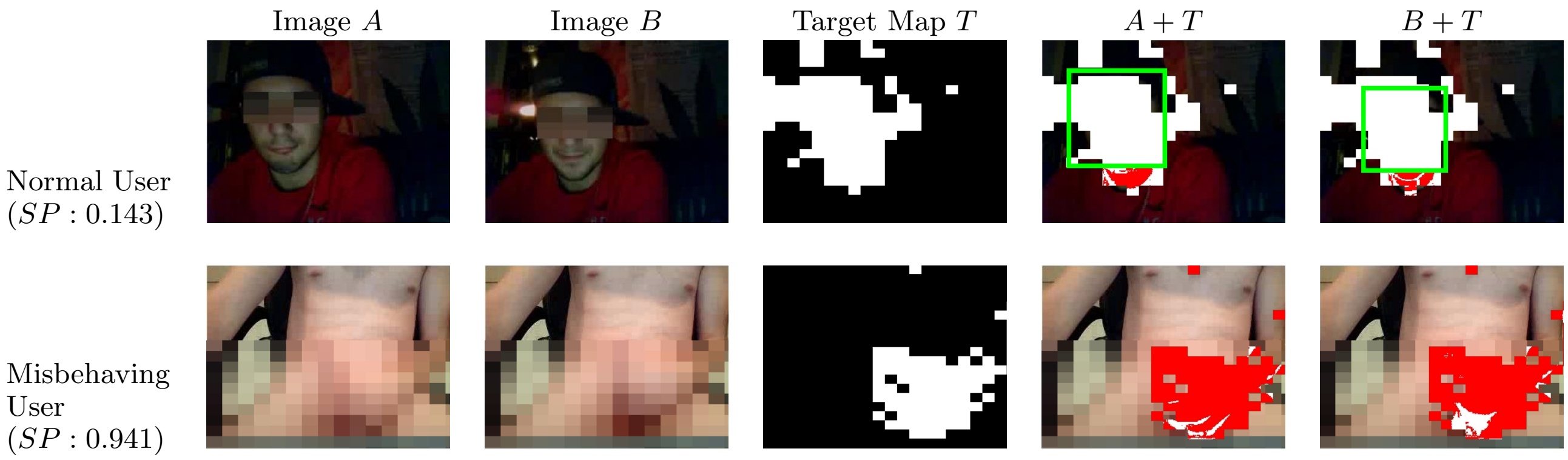

**Figure 4: Example of normal and misbehaving users. For each user, $A$ and $B$ are two screenshot images of the user; $T$ is the target map; $A+T$ and $B+T$ are original images overlaid with target region (white), non-face skin (red), and the detected face (green).** ***Skin proportion*** **($SP$) are 0.143 and 0.941 respectively, thus a good indicator for differentiating normal and misbehaving users.**

For each user, we can obtain multiple screenshots, resulting in multiple target maps. To select the best target map, we consider the size of the target region. Most video chat users keep a relatively stable pose and move only part of their body, such as head, hand or lower body. Thus a good target map should contain a target region that is *large enough to be a body part while most part of the map is stable or static.* Let $TA_{min}$ be the area of the smallest possible body part, which is derived from the training data and is set to 10% of image size in our experiments. We select the best target map for each user as follows: (a) if there are target maps with target region bigger than $TA_{min}$, select the one with the smallest target region; (b) if all target maps' target regions are smaller than $TA_{min}$, select the target map with the largest target region.

## 5.2 Skin Detector

Next, we detect exposed skin in the target region. Unlike typical pornographic or commercial images, skin color in online video chat images varies in a wide range due to diverse illumination conditions. Skin color does not always appear yellow or orange, and actually falls out of this range most of the time. Under different lighting, angle, and reflection from computer screen, skin color could appear pink, brown, blue, or even green in video chat images.

To address this problem, we consider three different skin color palettes. *Palette 1* is directly derived from the adaptive real-time skin detector in OpenCV, which could identify most yellow and orange skin. *Palette 2* adds the "pinkish" color to Palette 1 to cover most Caucasian skin colors under white lighting. This is achieved by converting RGB images to the HSV color space and detecting pixels with a Hue value in the range of $[0, 60]$ and $[300, 360]$. *Palette 3* focuses on the skin color of flashers, which usually has darker illumination. This is derived from a training dataset that contains only flashers with manually marked skin area.

Figure 5 shows the *skin proportion* ($SP$) (Section 5.3) of normal and misbehaving users under these three skin color palettes. An ideal skin palette should result in low $SP$ values for normal users and high $SP$ values for misbehaving users. As shown in the figure, Palette 1 performs well for normal users (shown in green); Palette 3 performs well for misbehaving users (shown in red); and Palette 2's performance is in the middle. Since no single palette is perfect, we propose an approach combining the three palettes, which will be discussed in Section 5.4.

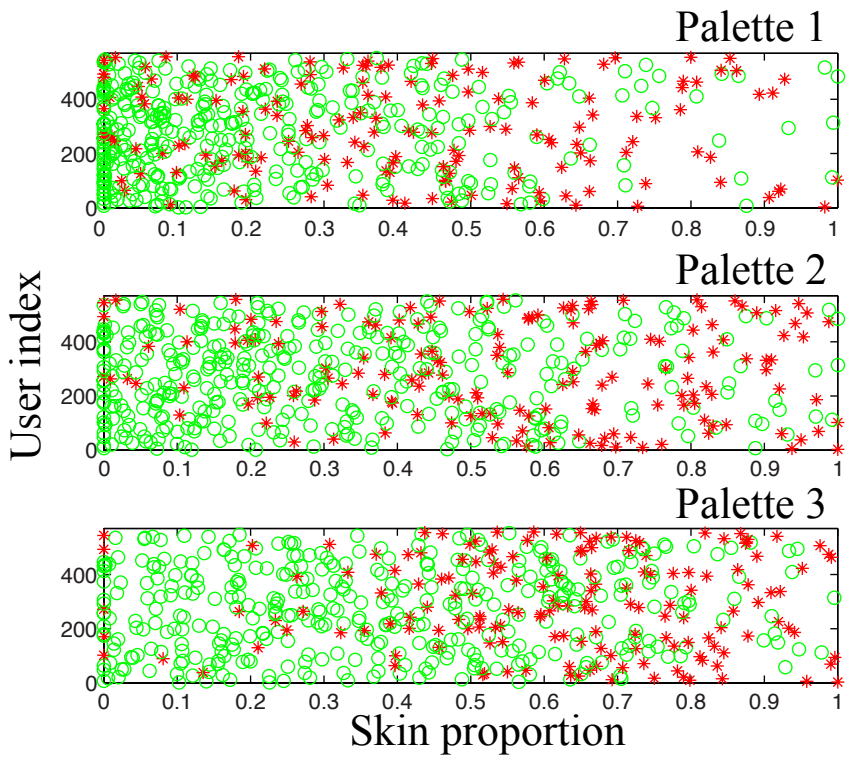

**Figure 5: Skin proportion of normal (green) and misbehaving (red) users under three different skin color palettes. Palette 1 performs well for normal users; Palette 3 performs well for misbehaving users; and Palette 2's performance is in the middle.**

## 5.3 Skin Proportion

To distinguish normal users from misbehaving users, we define the *Skin Proportion* ($SP$) as follows:

$$SP = \text{area of non-face skin}/\text{area of target region}. \quad (2)$$

Face skin area is determined using the frontal face detector in OpenCV. If no face is detected, all the detected skin is no-face skin. If there is a face, then non-face skin refers to all the skin below jaw of detected face. For each user, we select the best target map (Section 5.1), and compute two $SP$ values, one for each original image. Since a larger $SP$ value represents higher probability of being a flasher, to ensure that we identify all possible flashers, we choose the larger $SP$ value as the final $SP$ value for each user.

## 5.4 Misbehaving Probability

Let $SP(x)$ be the skin proportion value of a given user $x$, we need to determine the probability $p(x)$ of user $x$ being a misbehaving user. Since a user is either misbehaving or not, the distribution is binomial. A standard linear regression model which assumes normal distribution of the dependent variable is not appropriate. Instead, we use the binary logistic regression model, and the probability of success in the outcome (i.e., being a flasher) can be expressed as follows:

$$\log \frac{p(x)}{1 - p(x)} = \alpha + \beta \cdot SP(x). \tag{3}$$

As discussed in Section 5.2, we leverage three different measures (skin color palettes) to capture a user's skin exposure percentage (i.e., skin proportion). Therefore, the theoretical model above can be further expressed as

$$\log \frac{p(x)}{1 - p(x)} = \alpha + \beta_1 \cdot SP_1(x) + \beta_2 \cdot SP_2(x) + \beta_3 \cdot SP_3(x), \tag{4}$$

where $SP_1(x)$, $SP_2(x)$ and $SP_3(x)$ represent the skin proportion of user $x$ in the three different measures. Note that the three skin color palettes may be highly correlated and pose multicollinearity threats to our binary logistic regression model. To address this issue, we utilize Principal Component Analysis (PCA) to transfer multiple correlated variables into a smaller number of uncorrelated variables. Based on our experimental results (Section 7), we extract one component, called the *skin exposure component* ($SKC$), to represent the three aforementioned measures of skin exposure:

$$\log \frac{p(x)}{1 - p(x)} = \alpha + \beta \cdot SKC(x). \tag{5}$$

$SKC(x)$ is a linear function of normalized $SP$ scores $Z_{SP_i}$:

$$SKC(x) = a \cdot Z_{SP_1}(x) + b \cdot Z_{SP_2}(x) + c \cdot Z_{SP_3}(x) \tag{6}$$

where $Z_{SP_i}(x) = (SP_i(x) - avg(SP_i(X)))/stdev(SP_i(X))$. Here, $avg(SP_i(X))$ and $stdev(SP_i(X))$ are the average and standard deviation of the skin proportions of all users in a training dataset using the $i$-th skin color palette.

# 6. EVIDENCE FUSION

In addition to motion-based skin detection, we further boost system performance by considering different facial and upper body feature classifiers and combine all these classifiers using the Dempster-Shafer Theory of evidence. According to our observations described in Section 3, most misbehaving users do not show their faces. With further investigation, we find the presence of particular image characteristics like the face, eyes, nose, mouth, and upper body are all highly correlated with non-offensive content. Since individual classifiers look for different facial features of a face, their results could be seen as independent evidences of the face appearance. With our proposed classifier (i.e. motion-based skin detector) and all facial (i.e face, eye, nose, mouth) and upper body classifiers , we model these characteristics as a set of evidences that are used for obscene content detection. Each classifier is responsible for refining an individual evidence. In the following section, we will show how we harness these evidences to predict the likelihood of obscene content based on a few sequential screenshot images of a chatter.

## 6.1 Modeling in Dempster-Shafer Theory

Dempster-Shafer theory provides a representation of hypotheses through the definition of Belief function ($Bel$). The belief function is derived from a mass function, also called basic probability assignment ($m$). Assume that $\Theta$ represents all possible hypotheses and the basic probability assignment ($m$) is defined for every element $A$ of $2^\Theta$ ($A$ is also called the focal element), where $2^\Theta$ is the set of all subsets of $\Theta$. The value of the mass function $m(A) \in [0, 1]$ and

$$\begin{cases} m(\emptyset) = 0 \\ \sum_{A \in 2^\Theta} m(A) = 1 \end{cases} \tag{7}$$

where $\emptyset$ is the empty set. To illustrate this, we consider the example of misbehaving user detection. Assume a classifier is used for refining the evidence and the hypothesis space contains two hypotheses - a user is a flasher chatter ($H_F$) and a user is a normal chatter ($H_N$). We know that $2^\Theta = \{\emptyset, \{H_F\}, \{H_N\}, \{H_F, H_N\}\}$ and $m(\{H_F\}) + m(\{H_N\}) + m(\emptyset) + m(\{H_F, H_N\}) = 1$.

The belief function of hypothesis $A$, derived from basic probability assignment $m$, is defined as a mapping from $2^\Theta$ to the values in the interval $[0, 1]$

$$Bel(A) = \sum_{B \subseteq A} m(B) \tag{8}$$

In the Dempster-Shafer theory, the belief value of hypothesis $A$ is typically interpreted as the confidence level of the hypothesis. To illustrate how the belief function works, we continue with the misbehaving user detection example above. Suppose the evidence that the classifier refines can support hypothesis $H_F$ with 0.7 reliability and cannot support hypothesis $H_N$, i.e., the mass functions for set $\{H_F\}$ and $\{H_N\}$ are $m(\{H_F\}) = 0.7$ and $m(\{H_N\}) = 0$. Further, we can calculate the mass function for set $\{H_F, H_N\}$, i.e., $m(\{H_F, H_N\}) = 0.3$ using Equation 7. To obtain the confidence level of hypothesis $H_F$, we use Equation 8, i.e., $Bel(H_F) = 0.7$.

The definition of mass functions is dependent upon specific applications and is not provided by the Dempster-Shafer Theory. For image processing applications, the most widely used mass functions are derived either from probability at the pixel level [30] or from the distance to class centers [4]. In fault diagnosis applications, mass functions are defined based on subjective quantification [33]. Certainly, these quantifications are imprecise. In our application, we define mass functions as the performance of classifiers when the outcome of the classifiers are binary values; otherwise, mass functions are defined as detection probability. Recall that we harness some facial classifiers and an upper-body classifier to refine evidences from user's screenshot images, and the outcome of each classifier is a binary value. Since different binary classifiers have different classification precision, the reliability of a piece of evidence is significantly dependent upon the classification precision of the corresponding classifier. For example, binary classifier $C_i$ refines evidence $e_i$ and evidence $e_i$ supports hypothesis $H_j$. If evidence $e_i$ is 100% reliable and is present, hypothesis $H_j$ is true. However, binary classifier $C_i$ may make mistakes when refining evidence $e_i$, i.e., the precision is not equal to 1.0 when a binary classifier determines whether the evidence is present or not. Consequently, we use the precision of a binary classifier to define the corresponding mass functions. Our system also uses the motion-based skin detector and its outcomes are the probability that a user is a flasher as well as the probability that a user is a normal chatter. Therefore, we use these probabilities to define mass functions instead of using the classification precision.

In this application, obscene content detection is a binary classification; therefore, our hypothesis space only contains two hypotheses - a user is a normal chatter, and a user is a misbehaving chatter. Here we formulate these two hypotheses as $H_N$ and $H_F$ respectively. Furthermore, different evidences are used to support different hypotheses. Facial evidences (including the presence/absence of face, eyes, nose and mouth) and the upper-body evidence are only used for supporting hypothesis - $H_N$, and these evidences provide no supports for the opposite hypothesis - $H_F$. On the other hand, the skin exposure evidence can be used for supporting both hypothesis $H_N$ and hypothesis $H_F$. The main difference for these two types of evidences is that the beliefs of hypotheses that derive from the former evidences are non-additive and the beliefs of evidences that derive from the latter are additive because the motion-based skin detector is built upon probability theory while the other detectors are based on evidence theory [17].

Table 1: An example of evidence fusion.

| | $m_s(H_N) = 0.87$ | $m_s(H_F) = 0.13$ |
|---|---|---|
| $m_{f=1}(H_N) = 0.95$ | $\{H_N\} \to 0.8265$ | $\emptyset \to 0.1235$ |
| $m_{f=1}(H_N, H_F) = 0.05$ | $\{H_N\} \to 0.0435$ | $\{H_F\} \to 0.0065$ |

The following illustration gives an easy-to-understand introduction to the calculation of mass functions based on two different kinds of evidence - face evidence and skin evidence - which support two different hypotheses. Assume the face classifier's precision for correctly detecting the face evidence is 0.95 (i.e., when the face classifier identifies a face in a screenshot image, there is actually a face present in the image), and its precision for falsely detecting the face evidence is 0.32 (i.e., when the face classifier does not identify a face in a screenshot image, there is actually a face present in the image). Based on the definition of mass functions, we can calculate the mass functions based on face evidence as follows. (a) when the face classifier detects a face: $m_{f=1}(\{H_N\}) = 0.95$, $m_{f=1}(\{H_F\}) = 0$, and $m_{f=1}(\{H_N, H_F\}) = 0.05$; (b) when the classifier does not detect a face: $m_{f=0}(\{H_N\}) = 0.32$, $m_{f=0}(\{H_F\}) = 0$, and $m_{f=0}(\{H_N, H_F\}) = 0.68$.

Different from the mass function based on face evidence, the mass functions based on skin evidence are defined as follows. Assume the motion-based skin detector identifies a user as a normal chatter with 0.87 probability and as a misbehaving chatter with 0.13 probability. Then the mass functions based on skin exposure evidence are $m_s(\{H_N\}) = 0.87$, $m_s(\{H_F\}) = 0.13$ and $m_s(\{H_N, H_F\}) = 0$.

Based on the mass functions above, we further calculate the belief functions for hypothesis $H_N$ and $H_F$, and get the following results. When a face is detected by the face detector, the belief that a user is a normal chatter is 0.95 and the belief that a user is a misbehaving chatter is 0, i.e., $Bel(H_N) = 0.95$ and $Bel(H_F) = 0$. Conversely, the belief that a user is a normal chatter is 0.32 and the belief that a user is a misbehaving user is 0 when the face detector does not detect a face, i.e., $Bel(H_N) = 0.32$ and $Bel(H_F) = 0$. Apparently, these belief functions are non-additive (i.e., $Bel(H_N) + Bel(H_F) < 1$). On the other hand, when the skin detector indicates a user is a normal chatter with 0.87 probability and the user is a misbehaving chatter with 0.13 probability, the belief that a user is a normal chatter and the belief that a user is a misbehaving chatter are 0.87 and 0.13 respectively, i.e., $Bel(H_N) = 0.87$ and $Bel(H_F) = 0.13$. Notice that the belief functions are additive in this case.

## 6.2 Fusion and Decision Making

In this application, two hypotheses are supported by multiple evidences. Therefore, multiple evidences need to be combined in an effective way. The example in Section 6.1 indicates that hypothesis $H_N$ is supported by face evidence and skin exposure evidence. In order to obtain the belief of hypothesis $H_N$ from multiple evidences, we utilize a rule of combination that Shafer suggested [36], which allows mass functions to be combined in the following way:

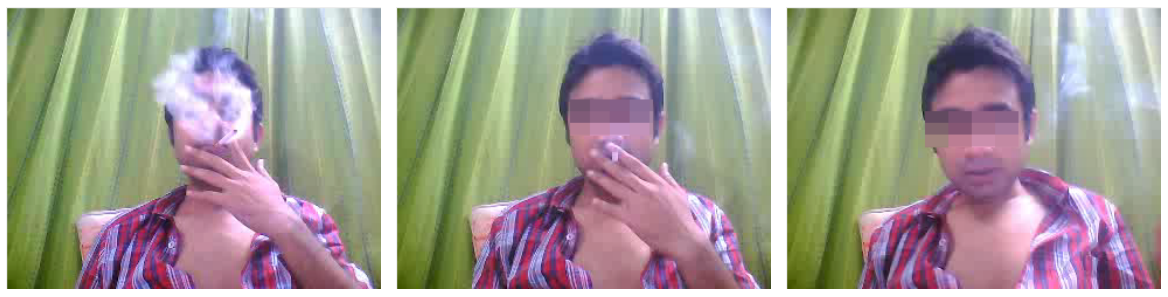

Figure 6: Sequential screenshot images of a user.

$$m_{i \bowtie j}(H) = \frac{\sum_{A \bigcap B = H \neq \emptyset} m_i(A) \cdot m_j(B)}{1 - \sum_{A \bigcap B = \emptyset} m_i(A) \cdot m_j(B)} \quad (9)$$

where $A, B, H \in 2^{\Theta}$ and $A \neq B \neq H$, $m_i$ and $m_j$ denote the mass functions based on evidence $i$ and evidence $j$ respectively. $i \bowtie j$ and $m_{i \bowtie j}$ represent the new combined evidence and the corresponding mass function based on the new evidence, respectively. To illustrate this, we continue the example introduced in Section 6.1 and assume that the face classifier identifies that a face is presented in a screenshot image. The combination results are summarized in Table 1. For each cell, we take the corresponding focal elements, intersect them and multiply their corresponding basic probability assignment. In the combined evidence, there are two focal elements - $\{H_N\}$ and $\{H_F\}$. The combined mass functions are calculated as follows:

$$m_{(f=1) \bowtie s}(\{H_N\}) = \frac{0.8265 + 0.0435}{1 - 0.1235} = 0.9926$$

$$m_{(f=1) \bowtie s}(\{H_F\}) = \frac{0.0065}{1 - 0.1235} = 0.0074$$

Based on the new combined evidence $(f = 1) \bowtie s$, we further calculate the belief of hypothesis $H_N$ and the belief of hypothesis $H_F$, which are $Bel(H_N) = 0.9926$ and $Bel(H_F) = 0.0074$. Thus, the user is more likely to be a normal chatter according to the combined evidence.

As introduced in Section 4, the system has the capacity of taking multiple screenshots for any user with a 10-second interval. This capacity allows our software to refine evidences in a more reliable way. In the current implementation, our software uses a sequence of three screenshot images from a user to make a decision. One of the advantages is that three sequential screenshot images of a user can reduce decision errors. Figure 6 shows three sequential screenshot images of a user. Our experiment indicates that the face detector cannot refine facial evidence from the first screenshot image because of the smoke, and the mouth detector cannot refine mouth evidence from the first two screenshot images because of the hand with a cigarette. However, the user's facial evidences are present in the third screenshot image. To address this issue, the decision making of our software uses the rule of maximum belief. Assume the belief values of hypothesis $H_N$, derived from three screenshot images, are $Bel_1(H_N)$, $Bel_2(H_N)$ and $Bel_3(H_N)$. The belief value of hypothesis $H_N$ for the user then is $Bel(H_N) = max\{Bel_1(H_N), Bel_2(H_N), Bel_3(H_N)\}$. Here, we use the maximum belief of hypothesis $H_N$ because it can greatly reduce the false positive rate and keep the false negative rate at a reasonable level.

# 7. EVALUATIONS

In this section, we evaluate the proposed solution for detecting obscene content and misbehaving users in online video chat services. We first describe the experimental setup and present the results of individual binary classifiers. We then focus on evaluating the performance of our motion-based skin detector and DST-based fusion technique.

## 7.1 Experimental Setup

As described in Section 3, screenshot images from 20,000 Chatroulette users are used in our experiments. We first filter out images that are dark or contain static scenes. Due to the limited space, details of this filtering process are omitted, and we refer interested readers to [2]. The categorization of the remaining 15,000 users are shown in Figure 1 (*filtered dataset*). These images are then randomly split into two subsets of equal size – one for training and the other one for testing. Since the binary classifiers (i.e., facial detectors and upper body detector) have already been well trained in the OpenCV library, we only use the training dataset to refine the reliabilities of these classifiers' evidences (the mass functions of these classifiers' evidences). Both the motion-based skin detector and DST-based fusion system are trained and tested using the training and testing datasets, respectively.

## 7.2 Binary Classifiers

Our system utilizes a number of binary classifiers from the OpenCV library to refine evidences. As described in Section 6, these classifiers do not give accurate detection, and a reliability value has to be assigned to each evidence that the corresponding classifier refines. Recall that we define the reliability value as the precision of the corresponding binary classifier. Intuition suggests that the performance of our fusion system is greatly dependent upon the reliability value assignment. Therefore, we randomly select 1,000 users' screenshot images from the training dataset to refine the reliability of each classifier, and place the selected screenshot images back to the training dataset (i.e., random selection with replacement). We repeat this operation $K$ times ($K = 10$) and use the mean value as the reliability value for the corresponding evidence. As shown in Table 2, the standard deviations of these evidences are fairly small, thus the mass functions which are used for the combination of facial and upper body evidences are relatively stable.

## 7.3 Motion-based Skin Detector

As described in Section 5, we consider three different skin color palettes and calculate the corresponding skin proportion $SP$ measures using motion-based target maps. The $SP$ measures are then combined in our binary logistic regression model to calculate the probability of a user being a flasher.

**Correlation Analysis and PCA.** Recall that we use PCA to transfer the three correlated $SP$ variables into one uncorrelated variable. Using the screenshot images in the training dataset, we first apply the method (described in Section 5) to calculate users' skin proportion $SP$ in three different measures and then performed a correlation analysis of the three $SP$ measures for each user. As shown in Table 3, the correlations are significant, and PCA is indeed needed in order to avoid multicollinearity threats to our regression model.

**Table 2: Mass Functions for Facial and Upper Body Evidences**

| $x$ | $m_{x=1}(H_N)$ | $m_{x=0}(H_N)$ | $stdev_{x=1}$ | $stdev_{x=0}$ |
|---|---|---|---|---|
| *face* | 0.984 | 0.327 | 0.017 | 0.018 |
| *eye* | 0.773 | 0.434 | 0.018 | 0.020 |
| *nose* | 0.802 | 0.455 | 0.029 | 0.030 |
| *mouth* | 0.711 | 0.219 | 0.016 | 0.020 |
| *upper body* | 0.821 | 0.491 | 0.030 | 0.025 |

**Table 3: Correlations among 3 Predictors**

| | $SP_1$ | $SP_2$ | $SP_3$ |
|---|---|---|---|
| $SP_1$ | 1 | 0.900** | 0.765** |
| $SP_2$ | 0.900** | 1 | 0.855** |
| $SP_3$ | 0.765** | 0.855** | 1 |

** Correlation is significant at the 0.01 level (2-tailed).

We used the PCA procedure in PASW18.0 (SPSS) [37] to transform the three measures of skin exposure. Kaiser Criterion (Eigenvalue > 1) was followed when selecting components and a scree plot (Figure 8) was used to confirm the dimensions identified. We extracted one component (skin exposure component $SKC$) to represent the three $SP$ measures according to the Eigenvalues and the elbow point identified in the scree plot. When subsequently building our binary logistic regression model, we only needed to consider $SKC$, which is a linear function of the normalized $SP$ scores:

$$SKC(x) = 0.362 \cdot Z_{SP_1}(x) + 0.384 \cdot Z_{SP_2}(x) + 0.349 \cdot Z_{SP_3}(x) \tag{10}$$

**Model Construction.** Since the training process of motion-based skin detector can be performed offline, we used the binary logistic regression procedure in the statistical package SYSTAT 13 [38]. Maximum Likelihood Estimation with EM algorithm was utilized in estimating the proposed model coefficients. The resulting regression model is as follows:

$$\log \frac{p(x)}{1 - p(x)} = -0.775 + 1.114 \cdot SKC(x) \tag{11}$$

Consider the Hosmer Lemeshow test [18], which is a test of goodness of fit for logistic regression model with continuous predictors, and a statistically non-significant test result ($\chi^2$) indicates a good fit of the model. Our proposed model provided a significantly good fit to the training data ($\chi^2 = 12.318, df = 8, p = 0.138$)[1] ($p < 0.05$). Specifically, our skin exposure composite is a statistically significant predictor for the probability of a user being a flasher. This was tested with Wald's test [40], which is a statistical test of significance for individual variable. The test statistics ($Wald = 43.108, df = 1, p = 0.000$) indicated that the influence of $SKC$ is statistically significant ($p < 0.05$).

**Model Performance.** Using the model constructed above,

[1] $df$ and $p$ denote degree of freedom and p-value, respectively.

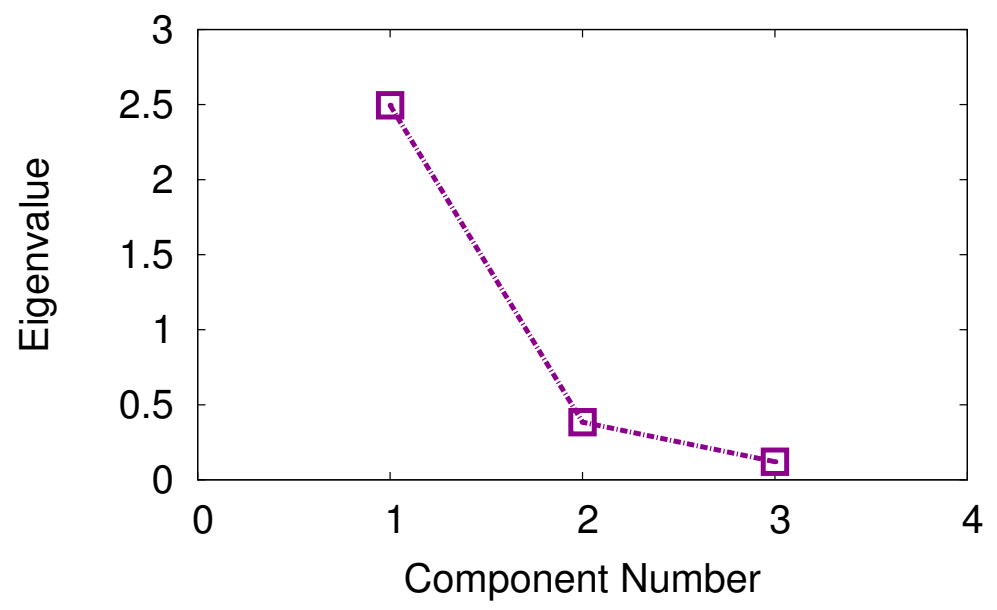

**Figure 8: Scree plot: Eigenvalues of extracted principal components.**

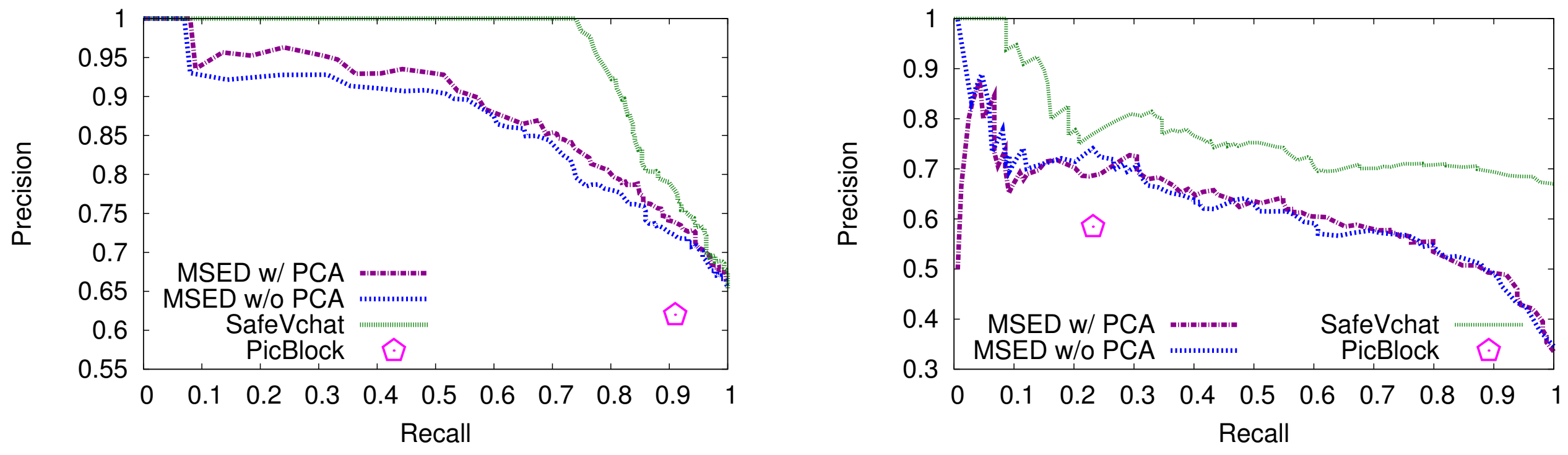

(a) Classification for normal users

(b) Classification for misbehaving users

**Figure 7: Performance comparison of obscene content detection. Our motion-based skin exposure detector (MSED) outperforms PicBlock, and our DST fusion based solution (SafeVchat) has the best performance.**

we then evaluated the model's performance in terms correctly classifying normal and misbehaving users. Figure 7 shows the precision-recall curves of different methods. As shown in Figure 7(a), when we only consider an individual skin color measure (no PCA), the performance for classifying normal users is slightly worse than that of the combination of three skin color measures using PCA. In contrast, Figure 7(b) shows that the performance for classifying misbehaving users remains approximately the same no matter whether PCA is used or not. To understand the reason behind this, we review the example in Figure 5. When we consider skin color in a small space, many misbehaving users are mixed with normal users. As the skin color space increases, a majority of misbehaving users are detected with more skin exposures and meanwhile a smaller number of normal users are also falsely detected. In Figure 5, there is an interesting observation – the normal users become identical and misbehaving users are still mixed with normal users when skin color space is enlarged. Therefore, a straightforward phenomenon shown in Figure 7 is that considering three measures improves the classification performance for normal users but not for misbehaving users.

## 7.4 Performance of SafeVchat

We now evaluate the performance of overall system, which uses DST to fuse the evidences of facial, upper body, and motion-based skin exposure detectors. The overall system is evaluated using the testing dataset, and the precision-recall curves for correctly classifying normal and misbehaving users are plotted in Figure 7. We can see that SafeVchat improves significantly in both precision and recall over other techniques.

We first compare SafeVchat with a state-of-the-art commercial software for detecting pornographic content – Pornographic Image Blocker (PicBlock 4.2.3). To the best of our knowledge, it uses both skin-color detection and text analysis techniques to determine whether an image on a website is objectionable. Its output is a binary value, i.e., either offensive or normal content. We used the same testing dataset to compare our system with PicBlock 4.2.3. As shown in Figure 7, the classification performance of PicBlock is much worse than our motion-based skin exposure detector and our DST-based fusion system. There are several reasons behind this. (1) There is no text information that can be used in online video chat systems. Therefore, the function of text analysis in PicBlock cannot help. (2) PicBlock does not consider motion between images, and may falsely identify some background as objectionable content. (3) Some misbehaving users do not show their entire body trunk. Instead, only a small proportion of skin region can be detected, which may bypass PicBlock's detection.

The fused system also outperforms our motion-based skin exposure detector. The reasons are the following. To reduce skin exposure detection errors, the skin exposure detector uses the face detector provided in the OpenCV library to crop the skin region on a face. However, the face detector cannot be operated in a perfect condition (e.g. a partial or side face cannot be detected). Therefore, some facial skin regions that cannot be identified by the face detector may be considered as large skin exposure, increasing the detection error in this condition. Using DST-based fusion, when the skin exposure detector mis-classifies a face-presented user, facial and upper body evidences, to some extent, can correct the classification errors and thus boost the performance of overall system.

We also observe in Figure 7 that the precision for classifying normal users remains high (almost 1.0) until the recall is above 0.75. The reason is that when there are positive facial or upper body evidences, the user is very likely to be a normal user (e.g., users who show their faces are very likely to be normal). However, when no such positive evidences are identified, more classification errors are possible (e.g., users who do not show their faces can be either normal or misbehaving), and the classification precision of the overall system approaches that of the skin exposure detector. On the other hand, the precision for classifying misbehaving users remains fairly stable (approximately 0.7) when the recall is in the range of $[0.5, 1.0]$. The reason is that when skin color space increases, more misbehaving users will be identified, meanwhile normal users who wear cloth with similar color to the skin color space may be misclassified as flashers. As we mentioned in Section 6, facial and upper body evidences can, to some extent, reduce the number of these misclassified users. Therefore, the classification precision remains relatively stable when the recall increases from 0.5 to 1.0.

# 8. CONCLUSIONS AND FUTURE WORK

This paper describes a system for detecting obscene content and identifying misbehaving users in online video chat services such as Chatroulette. A novel, motion-based skin detection method has been introduced. The results of individual binary classifiers such as face, eye, and nose detectors as well as our probabilistic motion-based skin detector are fused together into an overall decision classifying the user as misbehaving or not, based on Dempster-Shafer Theory. By using Chatroulette's real-world dataset, we have conducted a series of experiments, and evaluated our system performance. Compared to the current state-of-the-art software, PicBlock 4.2.3, our system achieves significantly higher classification precision and recall.

While our system has been specifically designed in the context of Chatroulette online video chat services, it can be extended to other webcam based online video chat systems. Our preliminary observations of several webcam based online chat rooms (such as Chat for Free [8], Goth Chat City [16] and iWebcam [1] etc.) show that the content transmitted and user behaviors in these online chat rooms are similar to Chatroulette. One difference is that the obscene content and behaviors are more common in such online chat rooms. We plan to experimentally explore how well our system performs in the context of online video chat rooms.

At present, we use three sequential screenshot images of a user to predict the likelihood that the user is a misbehaving user. It took on average 878 milliseconds to process three sequential screenshots of each user. While this response time is reasonable for obscene content detection, running our software and sequentially processing screenshot images on a large scale may result in high response times. In general, there are approximately 30,000 users online simultaneously. In the future, we will explore ways to scale our software. A software demo video is available in our project website [2].